\documentclass[aps,prl,superscriptaddress,floatfix,twocolumn]{revtex4}
\usepackage{graphicx}
\usepackage{amssymb}
\usepackage{float}
\usepackage[T1]{fontenc}
\usepackage[latin9]{inputenc}
\usepackage{amsbsy}
\usepackage{wrapfig}
\usepackage{caption}

\begin{document}

\title{Magnetically-Driven Suppression of Nematic Order in an Iron-Based Superconductor}

\author{S. Avci\footnote{Current address: Department of Materials Science and Engineering, 
Afyon Kocatepe University, 03200 Afyon, Turkey}}
\affiliation{Materials Science Division, Argonne National Laboratory, Argonne,
IL 60439-4845, USA}
\author{O. Chmaissem}
\affiliation{Materials Science Division, Argonne National Laboratory, Argonne,
IL 60439-4845, USA}
\affiliation{Physics Department, Northern Illinois University, DeKalb, IL 60115,
USA}
\author{J. M. Allred}
\author{S. Rosenkranz}
\affiliation{Materials Science Division, Argonne National Laboratory, Argonne,
IL 60439-4845, USA}
\author{I. Eremin}
\affiliation{Institut f\"ur Theoretische Physik III, Ruhr-Universit\"at Bochum, 
44801 Bochum, Germany}
\author{A. V. Chubukov}
\affiliation{Department of Physics, University of Wisconsin-Madison,
Madison,Wisconsin 53706, USA}
\author{D. E. Bugaris}
\author{D. Y. Chung}
\affiliation{Materials Science Division, Argonne National Laboratory, Argonne,
IL 60439-4845, USA}
\author{M. G. Kanatzidis}
\affiliation{Materials Science Division, Argonne National Laboratory, Argonne,
IL 60439-4845, USA}
\affiliation{Department of Chemistry, Northwestern University, Evanston, IL
60208-3113, USA}
\author{J.-P. Castellan}
\author{J. A. Schlueter}
\author{H. Claus}
\affiliation{Materials Science Division, Argonne National Laboratory, Argonne,
IL 60439-4845, USA}
\author{D. D. Khalyavin}
\author{P. Manuel}
\author{A. Daoud-Aladine}
\affiliation{ISIS Pulsed Neutron and Muon Facility, Rutherford Appleton
Laboratory, Chilton, Didcot OX11 0QX, United Kingdom}
\author{R. Osborn}
\affiliation{Materials Science Division, Argonne National Laboratory, Argonne,
IL 60439-4845, USA}
\email{ROsborn@anl.gov}

\begin{abstract}
A theory of superconductivity in the iron-based materials requires an
understanding of the phase diagram of the normal state. In these compounds,
superconductivity emerges when stripe spin density wave (SDW) order is
suppressed by doping, pressure or atomic disorder. This magnetic order is often
pre-empted by nematic order, whose origin is yet to be resolved. One scenario is
that nematic order is driven by orbital ordering of the iron 3$d$-electrons that
triggers stripe SDW order. Another is that magnetic interactions produce a
spin-nematic phase, which then induces orbital order. In this article, we report
the observation by neutron powder diffraction of an additional
four-fold-symmetric phase in Ba$_{1-x}$Na$_x$Fe$_2$As$_2$ close to the
suppression of SDW order, which is consistent with the predictions of
magnetically-driven models of nematic order. 
\end{abstract}

\date{\today}


\maketitle

\section{Introduction}
There have been extensive investigations of the phase diagrams of the various
iron arsenide and chalcogenide structures that display high temperature
superconductivity with critical temperatures up to 55
K~\cite{Stewart:2011kw,Paglione:2010ct,Johnston:2010cs,Canfield:2010ky}. In
common with other unconventional superconductors, such as the copper oxides,
heavy fermions, and organic charge-transfer salts, superconductivity is induced
by suppressing a magnetically ordered phase, which generates a high density of
magnetic fluctuations that could theoretically bind the Cooper pairs. Whether
this is responsible for the high transition temperatures has not been
conclusively established, but it makes the origin of the magnetic interactions
an important issue to be resolved~\cite{Dai:2012em,Lumsden:2010gt}.

In nearly all the iron arsenides and chalcogenides, the iron atoms form a square
planar net and the magnetic order consists of ferromagnetic stripes along one
iron-iron bond direction that are antiferromagnetically aligned along the
orthogonal iron-iron bond~\cite{Lumsden:2010gt,delaCruz:2008ej}. These systems
are metallic and the Fermi surfaces, which are formed by the iron 3$d$
electrons, are nearly cylindrical with hole pockets at the centre of the
Brillouin zone and electron pockets at the zone boundaries, all of similar size.
In such an electronic structure, interactions between electrons near the two
sets of pockets give rise to a spin density wave (SDW) order at the wavevector
connecting them~\cite{Eremin:2010ie}. This itinerant picture is consistent with
the wavevector of the observed antiferromagnetism, Angle Resolved Photoemission
(ARPES) measurements of the electronic structure~\cite{Ding:2008gi,Liu:2008hf},
and the evolution of the dynamic magnetic susceptibility with carrier
concentration~\cite{Castellan:2011en,Lee:2011hm,Luo:2012dc}.

However, any theory of the magnetic order also has to explain the structural
transition which occurs at a temperature either just above or coincident with
the SDW  transition and lowers the symmetry from tetragonal  ($C_4$) to
orthorhombic ($C_2$). This is often referred to as nematic order, and the
relation between nematicity, magnetic order, and superconductivity has become
one of the central questions in the iron-based
superconductors~\cite{Kasahara:2012ij,Allan:2013br}.

At present, there are two scenarios for the development of nematic order and
its relation to SDW order. In the first, the structural order is unrelated to
magnetism and is driven by orbital ordering as the primary instability.  The
orbital ordering induces magnetic anisotropy and triggers the magnetic
transition at a lower temperature by renormalizing the exchange
constants~\cite{Kruger:2009jm,Lv:2009bv,Chen:2010vc}.  This scenario is largely
phenomenological, but there have been recent efforts to develop a 
microscopic basis~\cite{Inoue:2012fb}.

In the second scenario, the structural order is driven by magnetic fluctuations,
associated with the fact that striped SDW order can be along the $x$-axis
(ordered momentum is $\mathbf{Q}_X = (0,\pi)$) or along the $y$-axis (ordered
momentum is $\mathbf{Q}_Y = (\pi,0)$). Theory predicts that the $Z_2$ symmetry
between the $X$ and $Y$ directions can be broken above the true SDW ordering
temperature that breaks $O(3)$ spin symmetry, {\it i.e.}, the system
distinguishes between $\mathbf{Q}_X$ and $\mathbf{Q}_Y$ without breaking
time-reversal symmetry~\cite{Fernandes:2012dv}. The order parameter of this
``Ising-spin-nematic state" couples linearly to the lattice, inducing both
structural and orbital order. The magnetic scenario has been developed for
itinerant~\cite{Eremin:2010ie,Fernandes:2012dv} and
localized~\cite{Chandra:1990ed,Xu:2008iy,Fang:2008by,Kamiya:2011gv} electrons,
and the phase diagrams are rather similar in the two approaches.  Below we use
the fact that the systems we study are metals and use an itinerant approach.

Many of the observable properties are identical in both the orbital and magnetic
scenarios, hindering a determination of the origin of nematicity. In the
following, we report the discovery of a new magnetic phase in hole-doped
Ba$_{1-x}$Na$_x$Fe$_2$As$_2$~\cite{Avci:2013iua,Aswartham:2012bw} at doping 
levels close to the suppression of magnetic order. This
second phase, which occurs at temperatures below the conventional $C_2$
transition, restores $C_4$ rotational symmetry, indicating that the SDW order
combines $\mathbf{Q}_X$ \textit{and} $\mathbf{Q}_Y$ with equal weights. Such a
second transition is highly unlikely in an orbital scenario because the breaking
of symmetry of $\mathbf{Q}_X$ and $\mathbf{Q}_Y$ is a pre-condition for a
magnetic transition to occur. However, it is known that such a phase is a
possible solution of itinerant magnetic models for certain  combinations of
electronic interactions and/or Fermi surface
geometries~\cite{Eremin:2010ie,Brydon:2011fd,Giovannetti:2011ed,Kang:2013vr}. By
going beyond our earlier Ginzburg-Landau analysis, we now show that the phase
diagram is much richer than previously thought and that the $C_4$ phase becomes
energetically favourable at higher doping levels, particularly in the range of
phase co-existence with superconductivity~\cite{Kang:2013vr}. We therefore view
the observation of the transition to an SDW state which does not break the
symmetry between $\mathbf{Q}_X$ and $\mathbf{Q}_Y$ as a strong indication that
the nematic order is of magnetic origin.

A magnetically-driven $C_4$ phase also provides a natural explanation for the new phase
observed in transport measurements in Ba$_{1-x}$K$_x$Fe$_2$As$_2$ under external
pressure~\cite{Hassinger:2012ki} and may explain anomalous diffraction results
in Ba(Fe$_{1-x}$Mn$_x$)$_2$As$_2$~\cite{Kim:2010cw}, so it is probably present
in other iron-based superconductors although our calculations show that its
stability is highly sensitive to details of the electronic structure.

In the following, we describe the experimental evidence for a reentrant $C_4$
phase in neutron and x-ray diffraction data on Ba$_{1-x}$Na$_x$Fe$_2$As$_2$ for 
$x>0.24$. We then summarize the results of theoretical calculations showing that such a phase 
is consistent with magnetically-driven nematic order.

\begin{figure}[!htb]
\centering
\includegraphics[width=0.9\columnwidth]{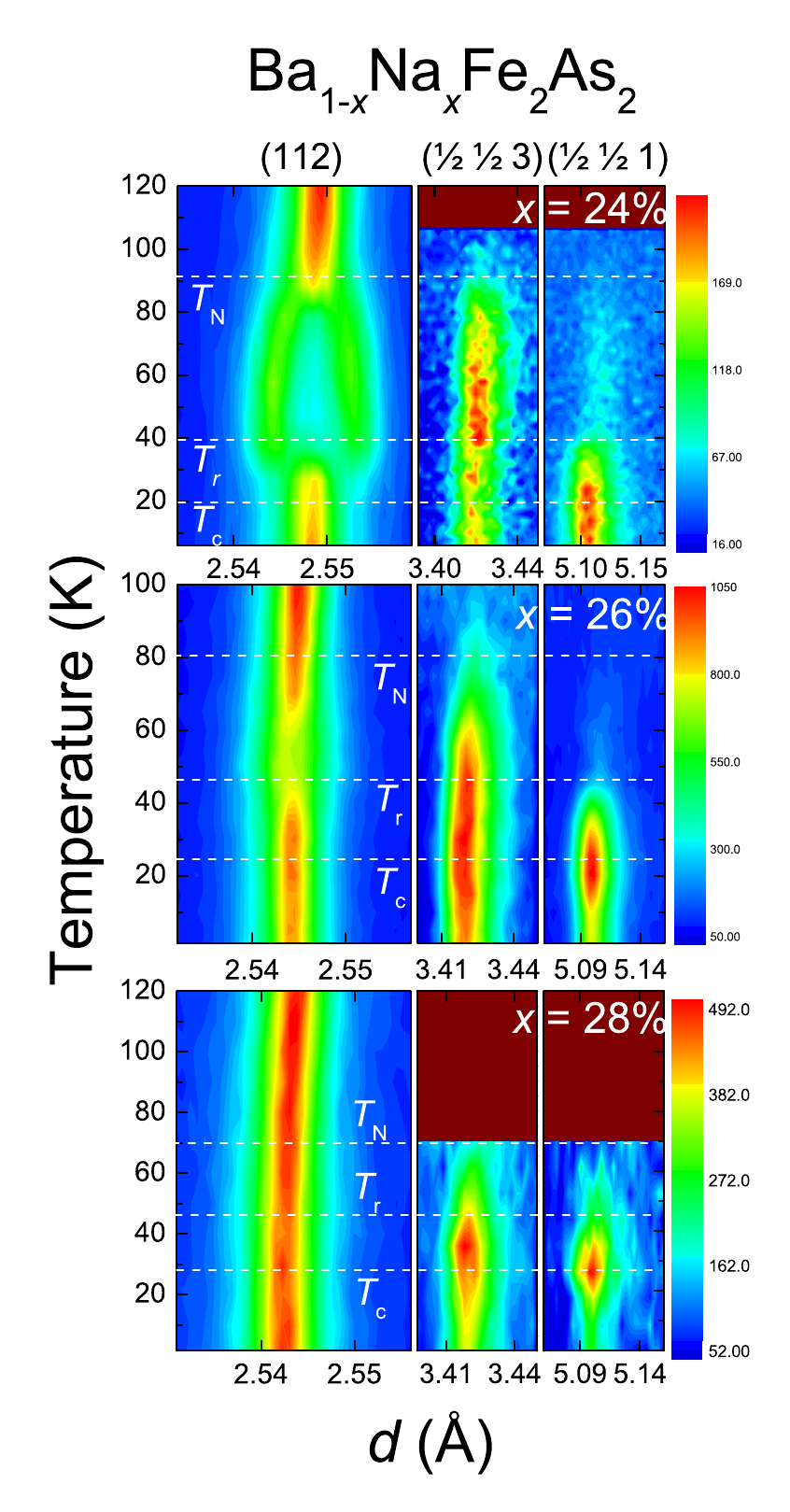}
\caption{\footnotesize
\textbf{Temperature dependence of powder neutron diffraction from
Ba$_{1-x}$Na$_{x}$Fe$_2$As$_2$} The first
diffractogram is of HRPD data from the (112) Bragg peak (using tetragonal
indices), which shows the orthorhombic transition at $T_{\rm N}$ and the reentrant
tetragonal transition at $T_{\rm r}$ in $x=0.24$ and 0.26. The symmetry is tetragonal
at all temperatures in $x=0.28$. The other two diffractograms are of Wish data
from magnetic Bragg peaks. The ($\frac{1}{2}\frac{1}{2}$3) data shows the onset
of stripe SDW order at $T_{\rm N}$. The ($\frac{1}{2}\frac{1}{2}$1) data shows the
onset of the $C_4$ SDW order at $T_{\rm r}$. The absolute intensities are arbitrary,
but to display all the plots on the same color scale, the
($\frac{1}{2}\frac{1}{2}$3) intensities have been multiplied by factors of 208,
200, and 144, and the ($\frac{1}{2}\frac{1}{2}$1) by factors of 30, 20, and 60,
for $x=0.24$, 0.26, and 0.28, respectively. The  magnetic Bragg peaks show a
significant reduction of intensity below the superconducting transition at
$T_{\rm c}$, indicating the phase competition between magnetism and superconductivity.
\label{DataFig} }
\end{figure}

\section{Results}
\subsection{Experiment}
We have conducted a detailed survey of the phase diagram of
Ba$_{1-x}$Na$_x$Fe$_2$As$_2$ using neutron and x-ray powder
diffraction~\cite{Avci:2013iua}, following our recent investigation of the
potassium-doped compounds~\cite{Avci:2012ha}. In both the K-doped and Na-doped
series, the addition of the alkali metal dopes holes into iron $d$-bands, and
reduces the transition temperature into the stripe phase from 139\,K, in the
parent compound BaFe$_2$As$_2$, to 0 at $x\sim 0.25-0.3$. One unusual feature of
both series is that the antiferromagnetic and orthorhombic transitions are
coincident and first-order over the entire phase diagram~\cite{Avci:2011cj}, an
observation that is quite unambiguous since both order parameters are determined
from the same neutron powder diffraction measurement. 

Details of the synthesis and characterization of the polycrystalline samples and the powder 
diffraction measurements are given in the Methods section. We provide a comparison 
of the sample stoichiometries with earlier reports in Supplementary Note 1.

The only region of the sodium series where there are significant departures from
the conventional behaviour observed in many iron-based superconductors is at
$0.24\leq x \leq 0.28$ close to the suppression of the AF/O order. These
compounds are all in the region where superconductivity coexists with magnetic
order at low temperature. The results are summarized in Fig.~\ref{DataFig},
where diffractograms are shown for three Bragg reflections at ($h$,$k$,$l$) =
(112), ($\frac{1}{2}\frac{1}{2}$1), and ($\frac{1}{2}\frac{1}{2}$3)
respectively, using tetragonal reciprocal lattice indices. The (112) reflection
is a nuclear Bragg peak that splits when the symmetry is lowered to
orthorhombic, while the other two reflections are magnetic Bragg peaks.

\begin{figure}[!t]
\centering
\includegraphics[width=\columnwidth]{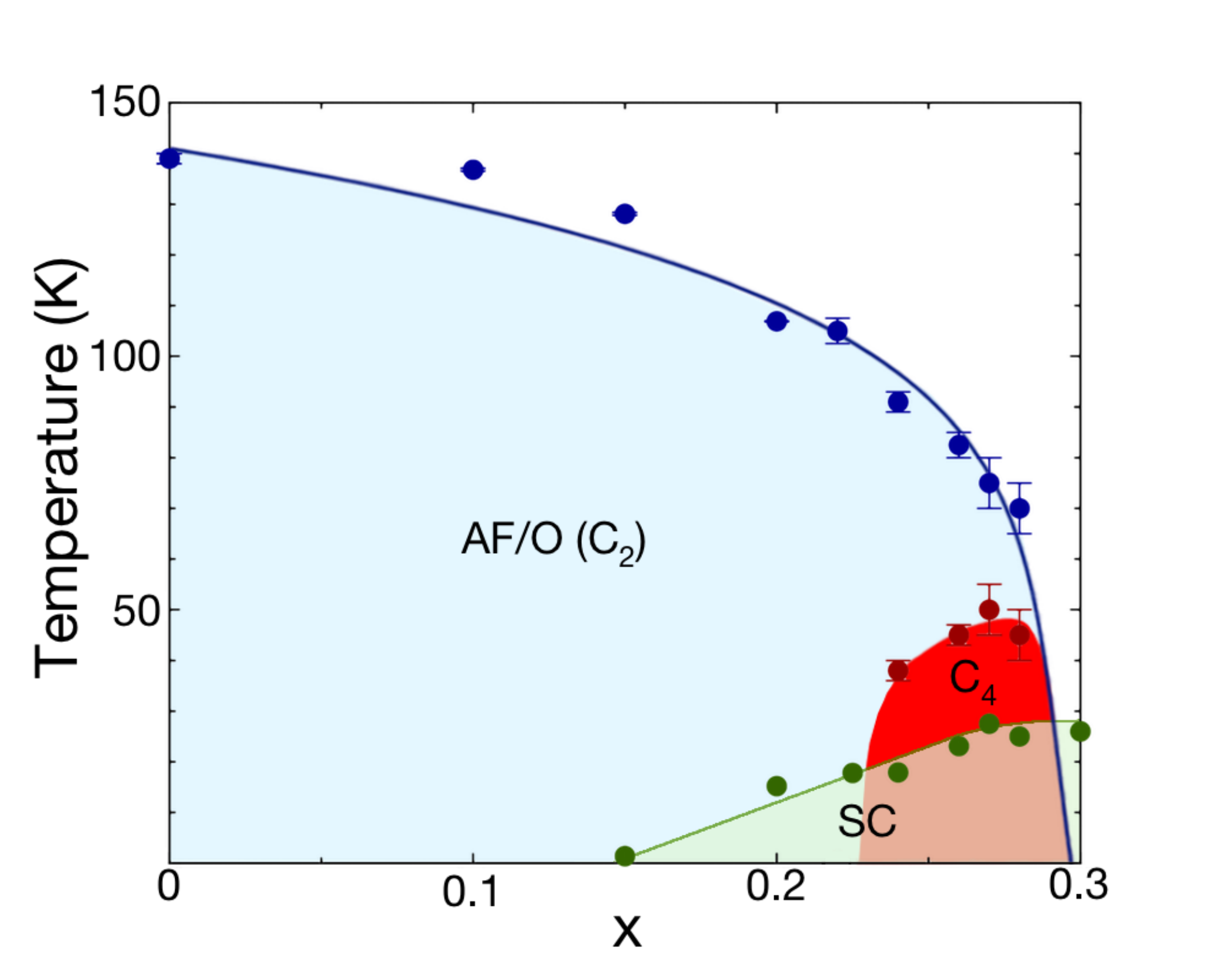}
\caption{\footnotesize
\textbf{Phase diagram of Ba$_{1-x}$Na$_x$Fe$_2$As$_2$} The blue points are the
coincident antiferromagnetic and orthorhombic transition temperatures, $T_{\rm N}$,
into the $C_2$ phase, and the red points are the observed transition
temperatures, $T_{\rm r}$ into the $C_4$ phase, all measured by neutron diffraction.
The green points are the superconducting transition temperatures, $T_{\rm c}$,
determined from magnetization data. The error bars represent the temperature interval in 
the neutron diffraction measurements.
\label{PhaseFig} }
\end{figure}

At $x =0.24$ and 0.26, the transition into the $C_2$ ($Fmmm$) phase at $T_{\rm N}\sim
70$-90~K is clearly evident. However, at $T_{\rm r}\sim40-50$\,K, there is a second
phase transition, not seen at $x=0.22$ (not shown), at which the orthorhombic
splitting collapses and tetragonal $C_4$ ($I4/mmm$) symmetry is restored. The
($\frac{1}{2}\frac{1}{2}$3) reflection, which shows the onset of stripe SDW
order at $T_{\rm N}$, weakens in intensity in the $C_4$ phase, whereas the
($\frac{1}{2}\frac{1}{2}$1) reflection strengthens considerably. This indicates
that there is a strong spin reorientation with respect to the stripe SDW order
when tetragonal symmetry is restored at $T_{\rm r}$. It was not possible to obtain an
unambiguous refinement of the $C_4$ magnetic structure so we cannot determine if
the reorientation is in-plane or out-of-plane. A full solution will require measurements on 
single crystals.

At $x=0.27$ (not shown) and 0.28, the temperature variation of the
($\frac{1}{2}\frac{1}{2}$3) and ($\frac{1}{2}\frac{1}{2}$1) reflections show
evidence of the same two magnetic transitions at $T_{\rm N}$ and $T_{\rm r}$, although the
orthorhombic splitting is too weak to be detected in the intermediate phase even
on a high-resolution diffractometer like HRPD.

These observations are summarized in the phase diagram of Fig.~\ref{PhaseFig},
which shows that the new phase is confined to doping levels very close to the
suppression of stripe SDW order. At $x=0.24$, the lower transition at $T_{\rm r}$ is
very sharp and appears to be first-order because there is evidence that up to
40\% of the sample remains in the $C_2$ phase below $T_{\rm r}$.  The $C_2$ phase
fraction is reduced to 20\% at $x=0.26$. It is not possible to determine if
there is phase coexistence at higher doping. Further details of the
coexistence of $C_2$ and $C_4$ phases at $x=0.24$ and 0.26 are 
provided in Supplementary Note 2.

Fig.~\ref{DataFig} shows that the $C_4$ phase competes with the
superconductivity because there is a strong suppression of the magnetic peak
intensities at temperatures below $T_{\rm c}$. This is similar to the phase
competition between superconductivity and the $C_2$ phase seen in the
electron-doped superconductors~\cite{Nandi:2010ea}, but much stronger than the
phase competition observed in the Ba$_{1-x}$K$_x$Fe$_2$As$_2$
series~\cite{Avci:2011cj}.

\subsection{Theory}

\begin{figure}[!htb]
\centering
\includegraphics[width=\columnwidth]{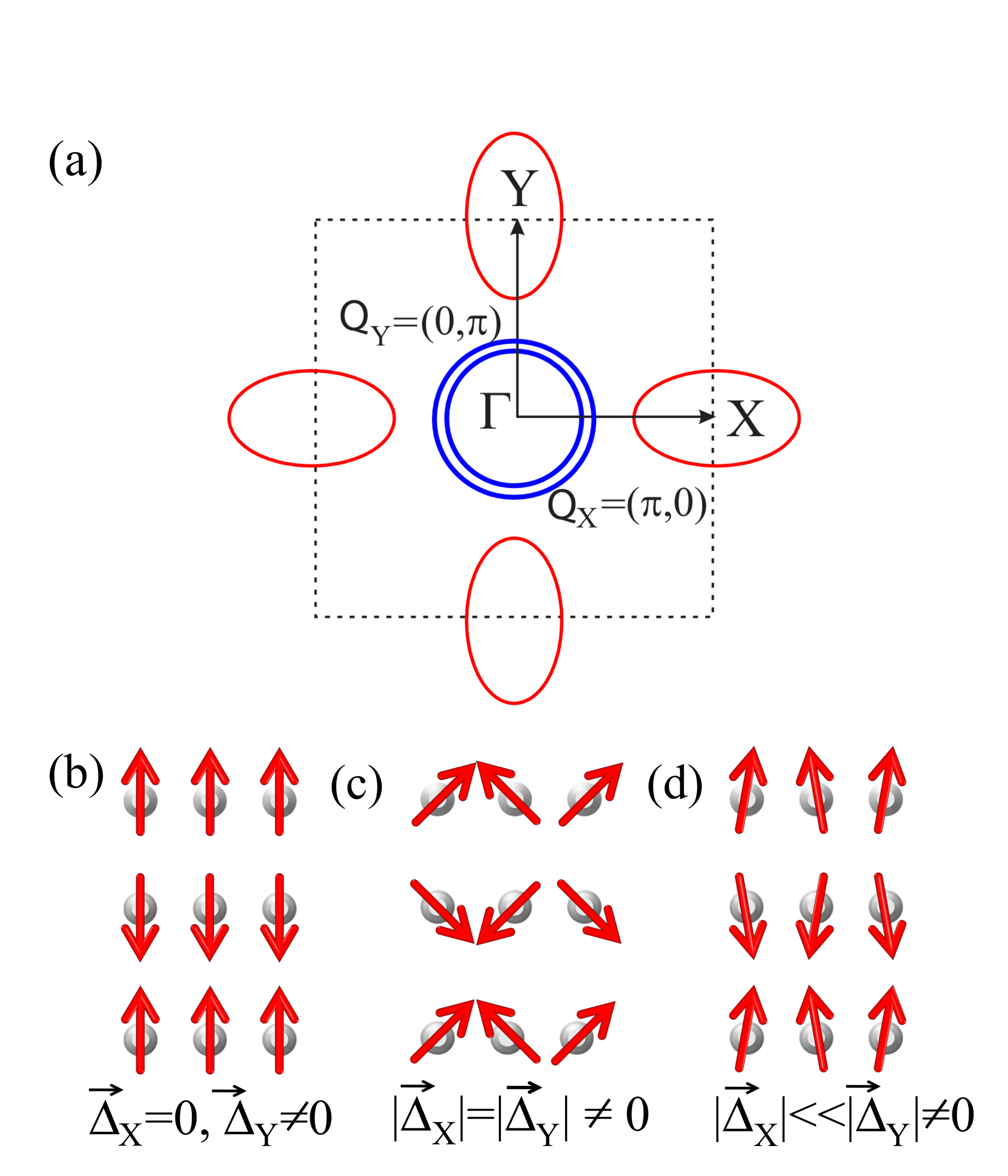}
\caption{\footnotesize
\textbf{Spin-nematic models of magnetic order} \textbf{(a)} The band-structure 
with two circular hole pockets at
$\Gamma$ and two electron pockets at X and Y, using the unfolded Brillouin zone
with one Fe atom per unit cell. The arrows refer to two equivalent nesting
wavevectors {\bf Q}$_X =(\pi,0)$ and ${\bf Q}_Y=(0,\pi)$. \textbf{(b,c,d)} Possible
magnetic ground states of the Fe-lattice:  (b) the $C_2$ antiferromagnetic
stripe phase with $\Delta_{X}=0$ and $\Delta_{Y} \neq 0$; (c) a $C_4$ magnetic
state, in which $|\Delta_{X}|=|\Delta_{Y}| \neq 0$, that is compatible with
tetragonal lattice symmetry (this is one of several possible solutions of the
 $C_4$ magnetic structures);  (d) magnetic order in which
$|\Delta_{X}|<<|\Delta_{Y}| \neq 0$. Note that the N\'eel transition at low
temperatures, $T_{\rm r}$, from phase (b) to phase (c) is first order, while the
transition from phase (b) to phase (d) is of the second order in which an
additional small component of $\Delta_{X}$ appears below $T_{\rm r}$. Our
experiments are more compatible with scenario (c).
\label{TheoryFig} }
\end{figure}

The itinerant description of magnetism in iron-based superconductors is built
upon the fact that the hole bands are centered around $\mathbf{Q}_\Gamma =
\left(0,0\right)$ and the electron bands are centered at
$\mathbf{Q}_X=\left(\pi,0\right)$ and $\mathbf{Q}_Y=\left(0,\pi\right)$,
respectively (Fig.~\ref{TheoryFig}(a)). The spin susceptibility is
logarithmically enhanced at momenta connecting the hole and electron pockets,
and SDW order develops even if the interaction is weak. The SDW order parameter
is in general a combination of the two vector components $\mathbf{\Delta}_X$ and
$\mathbf{\Delta}_Y$ with momenta $(\pi,0)$ and $(0,\pi)$, respectively. For a
model of  perfect Fermi surface nesting (circular hole and electron pockets of
equal radii) and only electron-hole interactions, SDW order determines
$|\mathbf{\Delta}_X|^2 + |\mathbf{\Delta}_Y|^2$ but not the relative magnitudes
and directions of  $\mathbf{\Delta}_X$ and $\mathbf{\Delta}_Y$.   Away from
perfect nesting,  the ellipticity of the electron pockets and interactions
between the electron bands break the degeneracy and lower the symmetry of the
SDW order. Near $T_{\rm N}$, an analysis within a Ginzburg-Landau expansion in powers
of $\mathbf{\Delta}_X$ and $\mathbf{\Delta}_Y$ shows that fourth-order terms
select stripe magnetic order with either $\mathbf{\Delta}_X \neq 0$,
$\mathbf{\Delta}_Y=0$, or $\mathbf{\Delta}_Y \neq 0$,
$\mathbf{\Delta}_X=0$~\cite{Eremin:2010ie,Fernandes:2012dv}. Such an order
simultaneously reduces the lattice $C_4$ symmetry down to $C_2$. The order
parameter in the stripe phase is shown schematically in Fig.~\ref{TheoryFig}(b).

An issue that has not been discussed in detail until now is whether another
magnetic ground state, in which both $\mathbf{\Delta}_X$ and $\mathbf{\Delta}_Y$
are non-zero, may appear at a lower temperature, as a result of non-linear
effects.  This might happen either \textit{via} a first-order transition, in
which case the most likely outcome is the state  in which $|\Delta_X| =
|\Delta_Y|$ (see Fig.~\ref{TheoryFig}(c)), or via a second-order transition, in
which case the second order parameter appears continuously and likely remains
relatively small down to $T=0$.

To check for a potential second SDW transition, we needed to go beyond the
previous Ginzburg-Landau analysis so we solved non-linear coupled mean-field
equations for $\Delta_X$ and  $\Delta_Y$ over the entire  temperature range and
analyzed which solution minimizes the free energy. This has revealed new
features in the phase diagram not previously identified. In particular, we  find
that SDW order with $\mathbf{\Delta}_X = \mathbf{\Delta}_Y$, which breaks $O(3)$
spin symmetry but preserves lattice $C_4$ symmetry, does emerge at low $T$, as
the mismatch in hole and electron pocket sizes grows.

We obtained this result by analyzing the minimal three-band model with  one hole
and two electron pockets. For simplicity, we considered parabolic dispersions
with
\begin{eqnarray}
\xi_{\Gamma,\mathbf{k}}&=&\varepsilon_{0}-\frac{k^{2}}{2m}-\mu\\
\xi_{X,\mathbf{k+Q_X}}&=&-\varepsilon_{0}+\frac{k_{x}^{2}}{2m_{x}}+\frac{k_{y}^{2}}{2m_{y}}-\mu\\
\xi_{Y,\mathbf{k+Q_Y}}&=&-\varepsilon_{0}+\frac{k_{x}^{2}}{2m_{y}}+\frac{k_{y}^{2}}{2m_{x}}-\mu
\end{eqnarray}
where $m_{i}$ are band masses, $\varepsilon_{0}$ is the offset energy, and $\mu$
is the chemical potential.

The non-interacting Hamiltonian takes the form
\begin{equation}
\mathcal{H}_{0}=\sum\limits _{i,\mathbf{k}}\xi_{i,\mathbf{k}}c_{i,\mathbf{k}\alpha}^{\dagger}c_{i,\mathbf{k}\alpha}\label{H_0}
\end{equation}
where $i=1-3$ label the bands, the summation over repeated spin indices $\alpha$
is assumed, and we shift the momenta of the fermions near the $X$ and $Y$ Fermi
pockets by $\mathbf{Q}_X$ and $\mathbf{Q}_Y$, respectively, writing
$\xi_{X,\mathbf{k+Q_X}}=\xi_{2,\mathbf{k}}$,
$\xi_{Y,\mathbf{k+Q_Y}}=\xi_{3,\mathbf{k}}$.

The interaction term in the Hamiltonian $\mathcal{H}_{int}$ contains all
symmetry-allowed interactions between low-energy fermions, which include inter-
and intra-band scattering processes~\cite{Chubukov:2008bp}. We present the
explicit form of $\mathcal{H}_{int}$ in the Supplementary Methods. The
mean-field equations on  $\mathbf{\Delta}_X$ and $\mathbf{\Delta}_Y$ are
obtained by introducing $\mathbf{\Delta}_X = (1/2N) \sum_k
c^\dagger_{1,\mathbf{k}\alpha} {\vec \sigma}_{\alpha \beta}  c_{2,
\mathbf{k}\beta}$ and $\mathbf{\Delta}_Y = (1/2N) \sum_k
c^\dagger_{1,\mathbf{k}\alpha} {\vec \sigma}_{\alpha \beta}  c_{3,
\mathbf{k}\beta}$ and using them to decouple four-fermion terms into anomalous
quadratic terms with inter-band ``hopping", which depends on $\mathbf{\Delta}_X$
and $\mathbf{\Delta}_Y$. We diagonalized the quadratic form, re-expressed $c_{i,
\mathbf{k}\alpha}$ in terms of new operators and obtained a set of two coupled
self-consistent equations for $\mathbf{\Delta}_X$ and $\mathbf{\Delta}_Y$.

We solved the mean-field equations numerically as a function of two parameters,
$\delta_0$ and $\delta_2$ (see Supplementary Methods for details). The
parameter $\delta_{0}=2\mu$ represents the mismatch in chemical potentials of
the hole and electron pockets ($\delta_{0}=0$ when the electron and hole pockets
are identical). $\delta_{2}=\varepsilon_{0}m(m_{x}-m_{y})/(2m_{x}m_{y})$ is
proportional to the ellipticity of the electron pockets. We focused on the two
SDW-ordered states-- the antiferromagnetic stripe state with $\mathbf{\Delta}_X
\neq 0$ and $\mathbf{\Delta}_Y=0$, in which $C_4$-symmetry is reduced to $C_2$
and on the SDW state with $\mathbf{\Delta}_X = \mathbf{\Delta}_Y$, in which
$C_4$-symmetry is preserved. As we said, the two states are degenerate at zero
ellipticity and perfect nesting, when  $\delta_2=\delta_0=0$. Once the
ellipticity becomes non-zero, the stripe state wins immediately below the N\'eel
temperature $T_{\rm N}$. The $C_4$-preserving state (with $\mathbf{\Delta}_X =
\mathbf{\Delta}_Y$) is a local maximum and is unstable at $T\leq T_{\rm N}$.

By solving the equations at lower temperature, we found that, at a finite
$\delta_0$, the $C_4$-preserving state also becomes locally stable below some 
$T<T_{\rm N}$, and at an even  lower $T<T_{\rm N}$, its free energy becomes smaller than
that of the stripe phase, \textit{i.e.}, at $T=T_{\rm r}$ the system undergoes a
first-order phase transition in which lattice $C_4$ symmetry gets restored. (see
Fig.~\ref{TheoryFig}(c)). Because $T_{\rm N}$ falls as the Fermi surface mismatch
$\delta_0$ increases, the new $C_4$-preserving phase in practice exists only in
a narrow region of the phase diagram close to the suppression of SDW order, as
observed in Fig.~\ref{PhaseFig}. We also analyzed a four-pocket model with two hole pockets
and found another scenario for a second SDW transition. Namely, the AF stripe
order $\mathbf{\Delta}_{Y}$ initially involves only fermions from a hole pocket
which has higher density of states. Below some $T<T_{\rm N}$, fermions near the
remaining hole pocket and near the electron pocket at $X$, not involved in the
initial stripe order, also produce SDW instability, and the system gradually
develops the second order parameter $|\mathbf{\Delta}_{X}|$, which distorts the
stripe AF order. The corresponding low $T$ spin configuration  is shown in 
Fig.~\ref{TheoryFig}(d). In this case, however,  the $C_4$ symmetry remains
broken at all temperatures. Our experimental data taken as a function of doping
are more consistent with a first-order transition and restoration of $C_4$
symmetry, although it is possible that the second scenario is realized under
pressure~\cite{Hassinger:2012ki}.

\section{Discussion}
We have demonstrated the existence of a wholly new magnetic phase that exists at
the boundary between superconductivity and stripe magnetism, an observation that
has important implications for the origin of magnetic and structural transitions
in the iron-based superconductors. It is important to distinguish these new
results from previous observations of a reentrant tetragonal phase in
electron-doped compounds, such as BaFe$_{2-x}$Co$_x$As$_2$~\cite{Nandi:2010ea}.
All those transitions were within the superconducting phase and have been shown
to result from the competition between superconductivity and stripe SDW
order~\cite{Fernandes:2010bn,Vorontsov:2010fv}. The reentrant phase that we
report here occurs at temperatures that are more than twice as high as $T_{\rm c}$ and
so requires a different explanation. However, there is a similar competition
between magnetism and superconductivity in the new phase evident from the
partial suppression of the ordered magnetic moment below $T_{\rm c}$.

We are unaware of any model of orbital order that would predict a reentrant
non-orbitally-ordered phase at lower temperature. However, the prediction of
spin-nematic models that a $C_4$ phase can become degenerate with the $C_2$
phase only at higher doping when the hole and electron Fermi surfaces are not as
well-matched in size, and that the stability of the $C_4$ phase would be limited
to a very narrow region close to the suppression of antiferromagnetism is borne
out by the new data.

Our results therefore provide strong evidence for the validity of an itinerant
model of nematic order in the iron-based superconductors, in which the orbital
reconstruction of the iron $3d$ states is a consequence of magnetic interactions
induced by Fermi surface nesting. Whether nematic order, or at least strong
nematic fluctuations, is a prerequisite for superconductivity is another
challenge to address in the future.

\section{Methods}
\subsection{Sample Synthesis}
Mixtures of Ba, Na, and FeAs were loaded in alumina tubes,
sealed in niobium tubes under argon, and sealed again in quartz tubes under
vacuum. The mixtures were variously subjected to 3 to 5 firings between 800 and
850$^\circ$C for 2-3 days for each firing, except for
Ba$_{0.78}$Na$_{0.22}$Fe$_2$As$_2$, which underwent two firings as above, and
then was heated for 16 hours at 1000$^\circ$C for each of the last two anneals.
Between each anneal, the powders were homogenized by grinding in a mortar and
pestle.  Annealing steps were kept as short as possible, enough to get
chemically homogeneous powders while minimizing sodium loss, which is
unavoidable. Before the last anneal, a slight amount of NaAs was added to
compensate for the loss. The structure and quality of the final black powders
were confirmed by x-ray powder-diffraction and magnetization measurements. The 
magnetization curves of the measured samples are shown in Supplementary Figure 1.
\subsection{Powder Diffraction}
The powder diffraction measurements were performed using two beam-lines at the
ISIS Pulsed Neutron Source, Rutherford Appleton Laboratory, UK: the high-resolution
powder diffractometer, HRPD, and the cold-neutron powder diffractometer, Wish. 
The high resolution available at HRPD was necessary in order to resolve the weak
orthorhombic splitting, while the high flux of Wish was required in order to
measure the weak magnetic reflections. The same samples were used on both
diffractometers within a few days of measurement. The results are summarized in
the diffractograms (plots of intensity \textit{vs d}-spacing and temperature),
shown in Fig. 1, with additional details provided by Supplementary Figure
2. 


\section{End Notes}
\subsection{Acknowledgements}

We are thankful to J. Knolle, R. Fernandes, J. Schmalian, R. Moessner, and V.
Stanev for useful discussions. This work was supported by the U.S. 
Department of Energy, Office of Science, Materials Sciences and Engineering 
Division (S.A., O.C.,  J.A., S.R., D.E.B., D.Y.C., M.G.K., J.-P.C., J.A.S., H.C., R.O.), which 
also supported A.V.C. under grant \#DE-FG02-ER46900 (A.V.C.).  I.E. acknowledges 
financial support from the DFG under priority programme SPP 1458 (ER 463/5) and 
the German Academic Exchange Service (DAAD PPP USA No. 57051534).

\subsection{Author contributions} The neutron and x-ray diffraction experiments
were devised by S.A., O.C., J.M.A., S.R., and R.O. and performed by S.A., O.C.,
J.M.A., and S.R. with experimental assistance from D.D.K., P.M., and A.D.A. The
Rietveld  refinements were performed by S.A. and O.C., with additional analysis
by D.D.K. and J.M.A. The theoretical calculations were performed by I.E. and
A.C. The samples were prepared by D.E.B., D.Y.C. and M.G.K. and characterized by
J.P.C., J.A.S., and H.C. The manuscript was written by S.A., O.C., J.M.A., I.E.,
A.C., and R.O., and the Supplementary Information was written by J.M.A., I.E. and A.C.

\subsection{Additional information} Supplementary Information is available in the
online version of the paper. Reprints and permissions information is available
online at www.nature.com/reprints. Correspondence and requests for materials
should be addressed to R. O. (Email: ROsborn@anl.gov).

\subsection{Competing financial interests}
The authors declare no competing financial interests.

\renewcommand{\thefigure}{S\arabic{figure}}
\setcounter{figure}{0}

\newpage
\section{Supplementary Notes}

\subsection{1. Sample Stoichiometry}
Using the synthesis conditions described in the article's Methods section, we 
observed well-defined superconducting
transitions (Supplementary Figure 1) in zero-field-cooled direct current (dc)
magnetization measurements at 0.02 Oe. The $x$ = 0.24, 0.27, and 0.28 samples
show sharp transitions similar to those reported in Ref~\cite{Avci13}. The $x =
0.26$ shows a somewhat broader transition although there is no evidence of
compositional fluctuations from neutron powder diffraction. In measurements
taken on the high resolution diffractometer, HRPD, the $x$ = 0.24, 0.26, and
0.28 samples all have sharp Bragg peaks. It is possible that the width of the
$x=0.26$ transition results from the intrinsic coexistence of orthorhombic and
tetragonal phases, which we discuss in the next section, but this needs further
investigation. The Bragg peak widths of the $x = 0.27$ sample are slightly
broader and the slight decrease in the magnetization of the $x=0.27$ sample at
34~K suggests some compositional inhomogeneity in this sample, although the
volume fraction would be too small to affect the neutron powder diffraction
results.

\captionsetup[figure]{labelformat=empty}
\begin{figure}[!htb]
\centering
\includegraphics[width=.8\columnwidth]{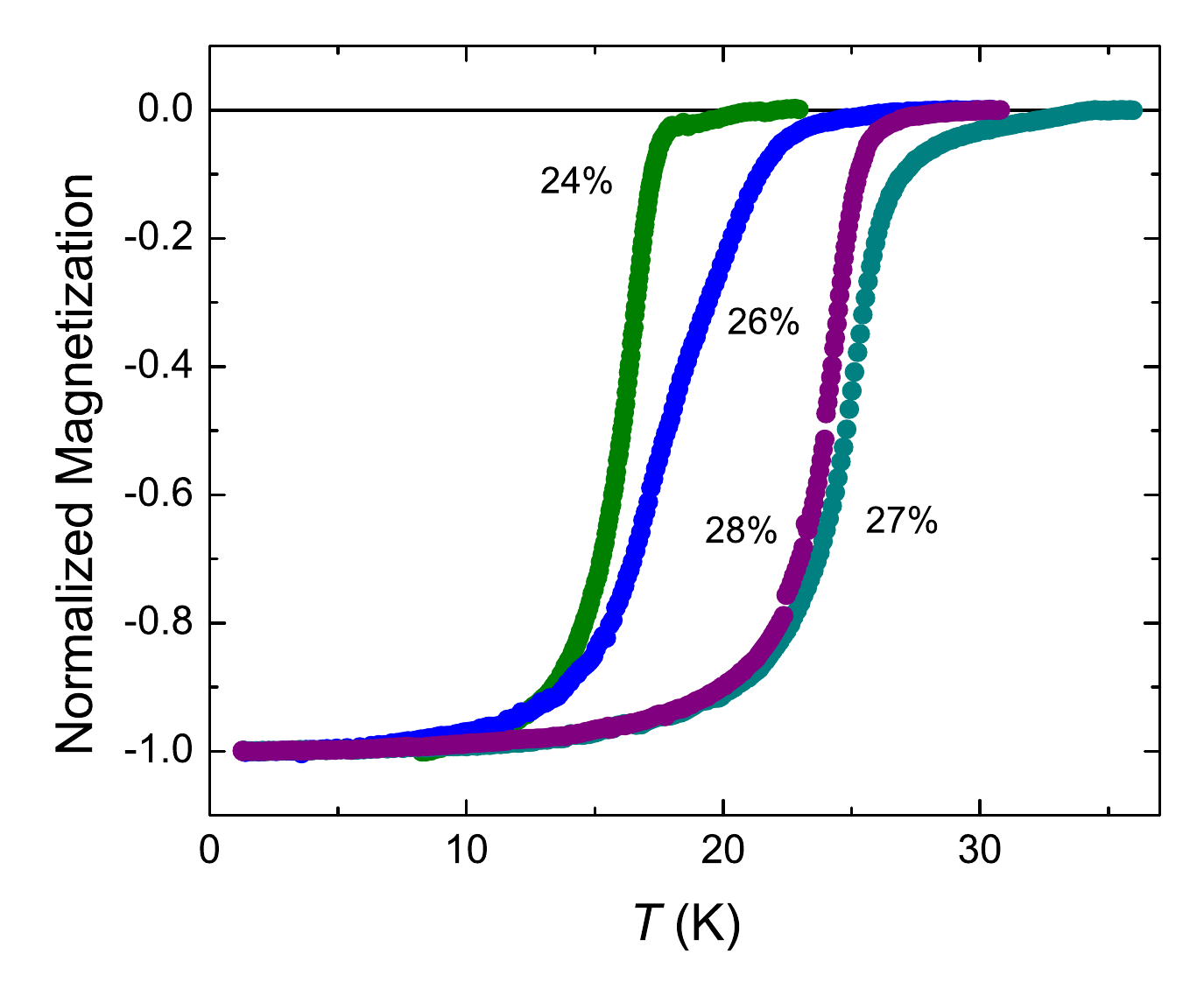}
\def\figurename{SUPPLEMENTARY FIGURE}
\caption{\footnotesize
DC magnetization for Ba$_{1-x}$Na$_x$Fe$_2$As$_2$ ($x$ = 0.24, 0.26,
0.27, and 0.28) in 0.02 Oe applied field.}
\label{FigS1}
\end{figure}

We note that Aswartham \textit{et al}~\cite{Aswartham12} show measurements on a
sample with a reported stoichiometry of $x=0.25$, without seeing any
thermodynamic anomalies associated with the $C_4$ transitions reported here.
However, they report $T_N=117$~K and $T_c=9$~K, which is more consistent with
$x\approx0.18$ in the phase diagram of Avci \textit{et al}~\cite{Avci13}, so
their sample is unlikely to be in the narrow compositional range in which we
observe the $C_4$ phase. However, as we mention in the Introduction, Hassinger 
 \textit{et al}~\cite{Hassinger12} see transport anomalies in Ba$_{1-x}$K$_x$Fe$_2$As$_2$ 
 under pressure that they attribute to a new spin-density-wave (SDW) phase. These
 would be consistent with our own observations if pressure stabilizes the $C_4$ phase
 at lower hole-dopings, which is reasonable since pressure also suppresses
 SDW order~\cite{Colombier09}.

\subsection{2. Determination of the Phase Diagram}

The temperature of the phase transition into the $C_2$ stripe phase,  $T_{\rm N}$, is
determined both by the onset of the orthorhombic splitting of the tetragonal
(112) peak and the intensity increase of  ($\frac{1}{2} \frac{1}{2}l$) magnetic
Bragg peaks, where $l=2n+1$. As in the other compositions reported by Avci
\textit{et al}~\cite{Avci13}, both transitions are coincident at $x=0.24$ and
0.26. However, the orthorhombic distortion becomes progressively weaker with
increasing $x$, so the splitting is only seen by the increase in peak widths at
$x=0.26$ (see the inset of Supplementary Figure2b) and cannot be resolved at all at
$x=0.27$ and 0.28. However, the $C_2$ transition is still evident in the
magnetic Bragg peak intensities, so these are used to determine  $T_{\rm N}$ at higher
doping. There is no evidence that the magnetic structure at $x=0.27$ and 0.28
differs from the stripe SDW order seen at lower doping, so these samples are
magnetically orthorhombic, even if the structural orthorhombicity is not
measurable.

\begin{figure}[!htb]
\centering
\includegraphics[width=.9\linewidth]{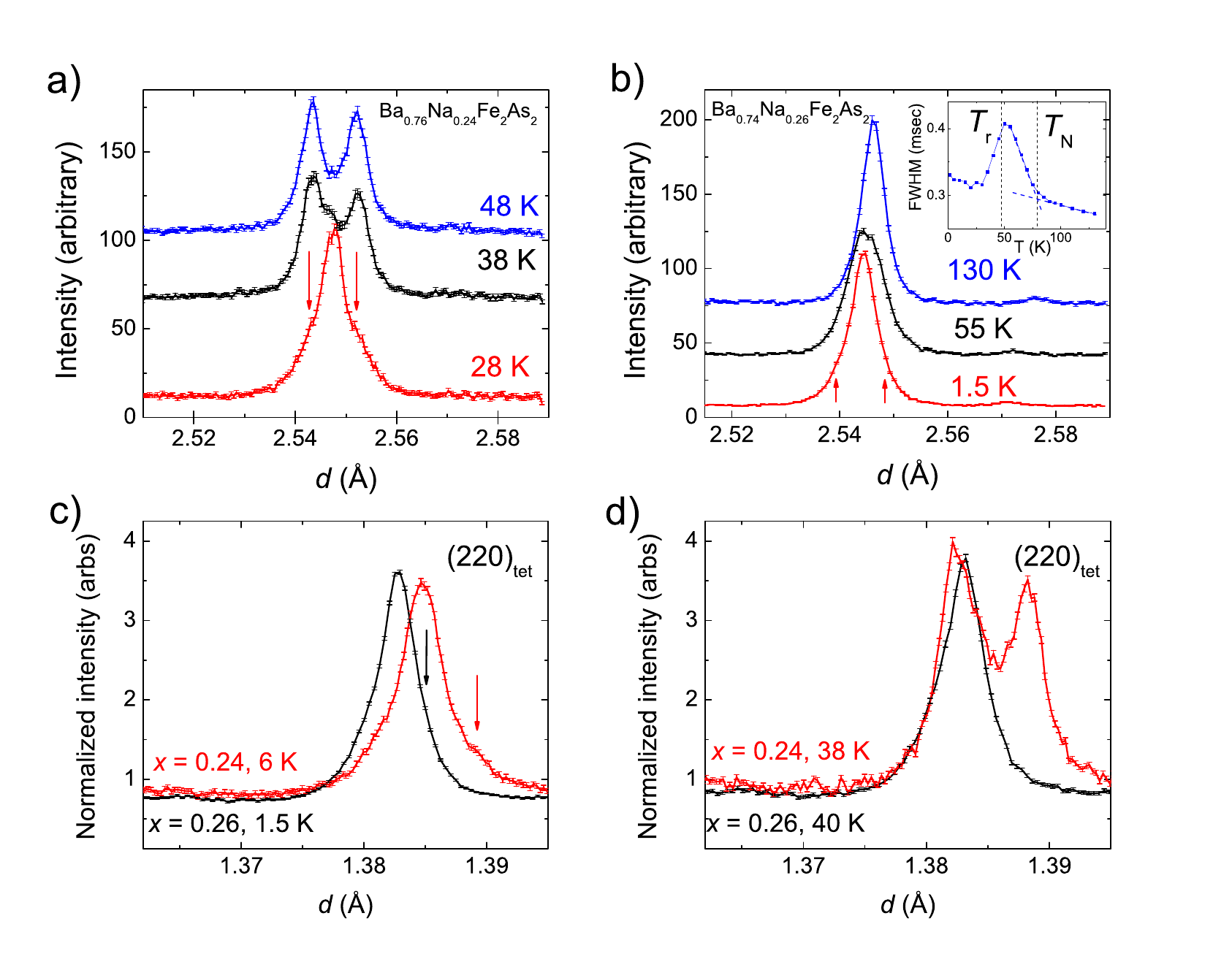}
\def\figurename{SUPPLEMENTARY FIGURE}
\caption{\footnotesize
(a) and (b): HRPD measurements of the (112) Bragg peak in
Ba$_{1-x}$Na$_x$Fe$_2$As$_2$ for a) $x = 0.24$ at $T = 28$~K, 38~K, and 48~K and
b) $x = 0.26$ at 1.5~K, 55~K, and 130~K. Arrows mark the position of the Bragg
peaks from the orthorhombic phase, which coexists with the tetragonal phase
below the $C_4$ transition.  Although the orthorhombic splitting cannot be seen
by eye in the $x=0.26$ sample, its onset is evident from the peak's full-width
at half maximum (inset) and gives rise to shoulders at 1.5~K that are not
present in the paramagnetic phase (130~K). (c) and (d): HRPD measurements of the
(220) Bragg peak in Ba$_{1-x}$Na$_x$Fe$_2$As$_2$ for $x = 0.24$ and 0.26 at c)
low temperature and d) $\sim40$~K. This comparison shows that there is little
overlap between the the $d$-spacings of the residual orthorhombic phases in both
samples. The error bars are derived from the square root of the raw detector counts.}
\label{FigS2}
\end{figure}

The diffractograms in Fig. 1 of the main article shows that the orthorhombic
splitting apparently collapses at the $C_4$ transition in $x=0.24$ and 0.26,
producing a reentrant tetragonal phase. However, there is really a phase
coexistence of the new tetragonal phase with a remnant orthorhombic phase, as
illustrated by the transfer of peak intensity from the orthorhombic peaks to the
tetragonal peaks shown in Supplementary Figure 2. The orthorhombic peaks persist with
approximately the same splitting down to the lowest temperatures, as can be seen
in the $x=0.24$ sample at 28~K in Supplementary Figure 2(a) and the $x=0.26$ sample at
1.5~K in Supplementary Figure 2(b). We estimate that approximately 40\% of the sample
remains orthorhombic at $x=0.24$ and 20\% at $x=0.26$. The coexistence of both
phases indicates that $C_4$ transition is first-order.

Since the orthorhombic splitting cannot be resolved at $x=0.27$ and 0.28, the
$C_4$ transition is most clearly seen in the rapid increase in the intensity of
the ($\frac{1}{2} \frac{1}{2}$1) peak seen in the WISH data. This peak is
present in the $C_2$ phase, but becomes much stronger in the $C_4$ phase,
reflecting a reorientation of the spins from the stripe phase. Therefore, we
have labelled the transition  $T_{\rm r}$, the 'r' stands for reorientation. Although
the peaks are too weak for a reliable Rietveld refinement of the magnetic
structure, the signature of the spin reorientation is the same in all four
samples so it is reasonable to assume that  $T_{\rm r}$ represents the phase boundary
from the $C_2$ phase into the $C_4$ phase over the entire range from $x=0.24$ to
0.28. It is not possible to say whether the transition remains first-order in
the higher-doped samples, since we cannot determine whether there is phase
coexistence in these compounds.

There are two scenarios that can explain the phase coexistence below  $T_{\rm r}$ in
some of the samples. One possibility is that it is due to heterogeneous
fluctuations in the local composition that straddle the phase line. However, if
this were the case, we would expect the remnant $C_2$ peaks within the $C_4$
phase to have a similar $d$-spacings to samples on the low-doped side of the
$C_4$ phase line, \textit{i.e.}, $x < 0.24$. However, Supplementary Figure 2(a) and
(b) shows that the residual $C_2$ shoulders are centered around the tetragonal
peak and that there is a large and unequivocal difference between the location
of these shoulders in the two samples. This is more clearly seen in Supplementary Figure 
2(c) and (d), where there is a significant difference between the
shoulders in $x = 0.26$ and the orthorhombic peaks in $x = 0.24$.  If the peak
broadening in $x = 0.26$ at low $T$ came from a fraction of the sample with
$x\lesssim0.24$ then the orthorhombic component would exhibit itself as
satellite peaks shifted away from the tetragonal peak instead of shoulders
surrounding it.

It is more likely that these samples are biphasic  where two phases of
equivalent composition coexist and the relative phase fractions are an
independent parameter. This is plausible because it implies that the energy
separation between ground states is very small, so that statistically both
phases must be present. Incomplete transformation could be a kinetic
effect\textemdash where the rate of cooling determines the relative phase
fractions\textemdash or it could be indicative of slow dynamic fluctuations
between orthorhombic and tetragonal symmetry.

\section{Supplementary Methods}

\subsection{Theory of the Reentrant $C_4$ Phase}

The magnetic phase in most parent compounds of the iron-based superconductors is
the stripe spin-density wave order with momentum either ${\bf Q_X}=(0,\pi)$ or
${\bf Q_Y}=(\pi,0)$ in the unfolded Brillouin zone, in which there is one iron
atom per unit cell~\cite{Yildirim08,Xiang08} (see Fig. 1(b) in the main
manuscript). These two wavevectors connect the hole pockets at the center of the
Brillouin zone and the electron pockets at the zone boundary, along the two
orthogonal iron-iron bond directions. The stripe magnetic ordering breaks both
$O(3)$ spin-rotational symmetry and $C_4$ lattice rotational symmetry (by
selecting either $\mathbf{Q}_X$ or $\mathbf{Q}_Y$),  and is often  preceded by a
``nematic'' phase in which $C_{4}$ symmetry is broken, but $O(3)$ rotational
symmetry remains unbroken.

In general, the geometry of iron pnictides allows more complex orders in which
both $\mathbf{\Delta}_X$ and $\mathbf{\Delta}_Y$ are present. Previous analysis
by two of us~\cite{Eremin10} have shown that near  $T_{\rm N}$, when Ginzburg-Landau
theory is valid, the system definitely prefers a stripe order in which only the
order parameter with $\mathbf{Q}_X$ or $\mathbf{Q}_Y$ is non-zero. Here we
extend the previous analysis to lower $T$ and study the magnetic order between
 $T=T_N$ and  $T=0$. Our results show that the phase diagram of the system is more
complex than previously thought. In particular, we find that at some range of
dopings, the system undergoes a first-order transition, upon lowering $T$,  into
a phase in which there is a two-component order parameter with equal magnitudes
of the components with $\mathbf{Q}_X$ and $\mathbf{Q}_Y$, and the four-fold
lattice rotational symmetry is restored (we label this phase as a $C_4$ phase).
A similar result has been recently reported in Ref.~\cite{kang}.  
The first order phase transition between the $C_2$ stripe and $C_4$  phases is
consistent with the observed by neutron diffraction experiments reported in the
main text.

We consider the minimal three-band model with the hole pocket $\Gamma$ at the
center of the Brillouin zone and two electron pockets $X$ and $Y$ at
$\mathbf{Q}_{X}$ and $\mathbf{Q}_{Y}$, respectively. For simplicity, we consider
parabolic dispersions with
$\xi_{\Gamma,\mathbf{k}}=\varepsilon_{0}-\frac{k^{2}}{2m}-\mu$,
$\xi_{X,\mathbf{k+Q_{X}}}=-\varepsilon_{0}+\frac{k_{x}^{2}}{2m_{x}}+\frac{k_{y}^
{2}}{2m_{y}}-\mu$, and  $\xi_{Y,\mathbf{k+Q_{Y}}}=-\varepsilon_{0}+\frac{k_{x}^{2}}{2m_{y}}+\frac{k_{y}^
{2}}{2m_{x}}-\mu$, where $m_{i}$ denotes the band masses, $\varepsilon_{0}$ is
the offset energy, and $\mu$ is the chemical potential. Near the Fermi energy
and for small ellipticity, the dispersions can be approximated by
$\xi_{\Gamma,\mathbf{k}}=-\xi$,
$\xi_{X,\mathbf{k+Q_{X}}}=\xi-\delta_{0}+\delta_{2}\cos2\theta$,
$\xi_{Y,\mathbf{k+Q_{Y}}}=\xi-\delta_{0}-\delta_{2}\cos2\theta$, with
$\delta_{0}=2\mu$, $\delta_{2}=\varepsilon_{0}m(m_{x}-m_{y})/(2m_{x}m_{y})$, and
$\theta=\tan^{-1}k_{y}/k_{x}$~\cite{Vorontsov10} .

Electrons with spin $\alpha$ of the band $i$ are created by the operators
$c_{i,\mathbf{k}\alpha}^{\dagger}$, and free-fermion part of the Hamiltonian has
the form
\begin{equation}
\mathcal{H}_{0}=\sum\limits _{i,\mathbf{k}}\xi_{i,\mathbf{k}}c_{i,\mathbf{k}\alpha}^{\dagger}c_{i,\mathbf{k}\alpha}\label{H_0}
\end{equation}
Here the summation over repeated spin indices is assumed, and we shift the
momenta of the fermions near the $X$ and $Y$ Fermi pockets by $\mathbf{Q}_{X}$
and $\mathbf{Q}_{Y}$, respectively, \textit{i.e.}, write
$\xi_{X,\mathbf{k+Q_{Y}}}=\xi_{X,\mathbf{k}}$,
$\xi_{Y,\mathbf{k+Q_{Y}}}=\xi_{Y,\mathbf{k}}$.

To shorten presentation, we restrict the interacting part of Hamiltonian to the
interaction in the spin channel with momenta near $\mathbf{Q}_{X}$ and
$\mathbf{Q}_{Y}$, i.e., to
\begin{equation}
\mathcal{H}_{\mathrm{int}}=-\frac{1}{2} U_{\mathrm{spin}}\sum\limits
_{i,\mathbf{q}}\mathbf{s}_{i,\mathbf{q}}\cdot\mathbf{s}_{i,-\mathbf{q}}\label{H_int}
\end{equation} 
where
$\mathbf{s}_{i,\mathbf{q}}=\sum_{k}c_{\Gamma,\mathbf{k+q}\alpha}^{\dagger}
\boldsymbol{\sigma}_{\alpha\beta}c_{i,\mathbf{k}\beta}$ 
is the electronic spin
operator, and $\boldsymbol{\sigma}_{\alpha\beta}$ are Pauli matrices. The
coupling $U_{\mathrm{spin}}$ is the sum of density-density and pair-hopping
interactions between hole and electron states ($U_{\mathrm{spin}} = U_{1} +
U_{3}$ in the notation of Ref. \cite{Chubukov08}), where
\begin{eqnarray}
U_{1}c_{\Gamma,\alpha}^{\dagger}c_{\Gamma,\alpha}c_{X,\beta}^{\dagger}c_{X,\beta} & = & -\frac{U_{1}}{2}c_{\Gamma,\alpha}^{\dagger}\boldsymbol{\sigma}_{\alpha\beta}c_{X,\beta}\cdot c_{X,\gamma}^{\dagger}\boldsymbol{\sigma}_{\gamma\delta}c_{\Gamma,\delta}\nonumber \\
 &  & +(\cdots)\nonumber \\
U_{3}c_{\Gamma,\alpha}^{\dagger}c_{X,\alpha}c_{\Gamma,\beta}^{\dagger}c_{X,\beta} & = & -\frac{U_{3}}{2}c_{\Gamma,\alpha}^{\dagger}\boldsymbol{\sigma}_{\alpha\beta}c_{X,\beta}\cdot c_{\Gamma,\gamma}^{\dagger}\boldsymbol{\sigma}_{\gamma\delta}c_{X,\delta}\nonumber \\
 &  & +(\cdots)\label{SDW_channel}
 \end{eqnarray}
and the dots stand for the terms with
$\delta_{\alpha,\beta}\delta_{\gamma,\delta}$, which only contribute to the CDW
channel. The couplings $U_1$ and $U_3$ do depend on the angle along the electron
pockets~\cite{Vishwanath09}, but for our purposes this dependence may be
neglected, i.e., $U_{\mathrm{spin}}$ can be approximated by a constant. Once
$U_{\mathrm{spin}}$ exceeds some critical value (which gets larger when
$\delta_{0}$ and $\delta_{2}$ increase), the static magnetic susceptibility
diverges at $(0,\pi)$ and $(\pi,0)$, and the system develops long-range magnetic
order.

To understand what kind of magnetic order wins below  $T_{\rm N}$ we introduce the two
spin  fields $\boldsymbol{\Delta}_{(X,Y)} = U_{spin}
\sum_{\mathbf{k}}c_{\Gamma,\mathbf{k}\alpha}^{\dagger}\boldsymbol{\sigma}_{\alpha\beta}c_{(X,Y),\mathbf{k}\beta}$. 
We apply Hubbard-Stratonovich transformation, integrate out fermions, obtain the 
action $S[\boldsymbol{\Delta}_{X}, \boldsymbol{\Delta}_{Y}$] in terms of
$\boldsymbol{\Delta}_{X}$ and $\boldsymbol{\Delta}_{Y}$, use saddle-point
approximation $\partial S/\partial {\boldsymbol{\Delta}}_i =0$, and solve a set
of coupled saddle-point equations for $\boldsymbol{\Delta}_{X}$ and
$\boldsymbol{\Delta}_{Y}$  A straightforward way to perform this calculation is 
to introduce the $6$-dimensional Nambu operator:
\begin{equation}
\Psi_{\mathbf{k}}^{\dagger}=\left(\begin{array}{cccccc}
c_{\Gamma,\mathbf{k}\uparrow}^{\dagger} & c_{\Gamma,\mathbf{k}\downarrow}^{\dagger} & c_{X,\mathbf{k}\uparrow}^{\dagger} & c_{X,\mathbf{k}\downarrow}^{\dagger} & c_{Y,\mathbf{k}\uparrow}^{\dagger} & c_{Y,\mathbf{k}\downarrow}^{\dagger}\end{array}\right)\label{A_nambu}
\end{equation}
Applying the Hubbard-Stratonovich transformation and evaluating the products of
the Pauli matrices, we obtain the partition function in the
form~\cite{Fernandes10,Fernandes12}:
\begin{equation}
Z=\int d\Delta_{i}d\Psi\mathrm{e}^{-S\left[\Psi,\Delta_{i}\right]}\label{A_Z}
\end{equation}
with the action
\begin{equation}
S\left[\Psi,\Delta_{i}\right]=-\int_{k}\Psi_{k}^{\dagger}\mathcal{G}_{k}^{-1}\Psi_{k}^{}+\frac{2}{U_{\mathrm{spin}}}\int_{x}\left(\Delta_{X}^{2}+\Delta_{Y}^{2}\right)\label{A_S}
\end{equation}
Here $\Delta_{i}=\left|\boldsymbol{\Delta}_{i}\right|$, and the Green's function
$\mathcal{G}_{k}^{-1}$ is given by:
\begin{equation}
\mathcal{G}_{k}^{-1}=\mathcal{G}_{0,k}^{-1}-\mathcal{V}\label{A_G}
\end{equation}
where the free-fermion term is
\begin{equation}
\mathcal{G}_{0,k}=\left(\begin{array}{ccc}
\hat{G}_{\Gamma,k} & 0 & 0\\
0 & \hat{G}_{X,k} & 0\\
0 & 0 & \hat{G}_{Y,k}\end{array}\right)\label{A_G0}
\end{equation}
and the interaction term is
\begin{equation}
\mathcal{V}=\left(\begin{array}{ccc}
0 & -\hat{\Delta}_{X} & -\hat{\Delta}_{Y}\\
-\hat{\Delta}_{X} & 0 & 0\\
-\hat{\Delta}_{Y} & 0 & 0\end{array}\right)\label{A_V}
\end{equation}
Here we introduced the $2\times2$ matrices $\hat{G}_{i,k}=G_{i,k}\mathbb{I}$ and
$\hat{\Delta}_{i}=\boldsymbol{\Delta}_{i}\cdot\boldsymbol{\sigma}$, where
$\mathbb{I}$ is the identity matrix. The functions
$G_{i,k}^{-1}=i\nu_{n}-\xi_{i,\mathbf{k}}$ are the non-interacting
single-particle Green's functions for $\Gamma$, Y, and X fermions.

It is now straightforward to integrate out the fermions, since the action is
quadratic in them, and obtain the effective magnetic action:
{\footnotesize
\begin{equation}
S_{\mathrm{eff}}\left[\boldsymbol{\Delta}_{X},\boldsymbol{\Delta}_{Y}\right]=-\mathrm{Tr}\ln\left(1-\mathcal{G}_{0,k}\mathcal{V}\right)+\frac{2}{u_{\mathrm{spin}}}\int_{x}\left(\Delta_{X}^{2}+\Delta_{Y}^{2}\right)\label{A_aux_effective_S1}
\end{equation}}

Here $\mathrm{Tr}\left(\cdots\right)$ refers to the sum over momentum, frequency
and Nambu indices. In contrast to the previous
studies~\cite{Fernandes12,Eremin10}, we do not perform a series expansion in
powers of $\Delta_{i}^{2}$ but analyze the full non-linear saddle-point
(mean-field) solutions $\frac{\delta S}{ \delta \Delta_{i}} =0$ for the two
cases: $(i)$ a $C_2$ stripe phase, in which we set $\Delta_{X}\neq 0$,
$\Delta_{Y}=0$ and $(ii)$ a $C_4$ phase in which we set $\Delta_{X} = \Delta_{Y}
= \Delta$ For the case $(i)$, the mean-field equation has the form
\begin{equation}
1= 2 U_{spin} \Delta_X \sum_{{\bf k}, i\nu_n} \frac{1}{\Delta^2_X-G^{-1}_{\Gamma,k}G^{-1}_{X,k}}
\end{equation}
while for the case $(ii)$ we have
\begin{equation}
1= 2 U_{spin} \Delta \sum_{{\bf k}, i\nu_n} \frac{1}{\Delta^2-G^{-1}_{\Gamma, k}G^{-1}_{X, k}+\Delta^2 G^{-1}_{X, k}G_{Y, k}}
\end{equation}
In both cases the sum over Matsubara frequencies can be evaluated exactly. For
stripe magnetic order, the mean-field equation becomes
\begin{equation}
1= - 2 U_{spin} \Delta_X \sum_{{\bf k}} \frac{f\left(E_{1{\bf k}}\right)-f\left(E_{2{\bf k}}\right)}{E_{1{\bf k}}-E_{2{\bf k}}}
\end{equation}
where 

{\footnotesize $E_{1,2{\bf k}}= \frac{1}{2} \left(\xi_{\Gamma,{\bf k}}+\xi_{X,{\bf
k+Q_X}} \pm \sqrt{\left(\xi_{\Gamma,{\bf k}}-\xi_{X,{\bf k+Q_X}}\right)^2+4
\Delta_X^2} \right)$.} 

For the $C_4$ phase, we obtain
\begin{widetext}
\begin{eqnarray}
1 = - 2 U_{spin} \Delta \sum_{{\bf k}} && \left[ \frac{\left(E_{11{\bf k}}-\xi_{X,{\bf k+Q_Y}}\right)f\left(E_{11{\bf k}}\right)}{\left(E_{11{\bf k}}-E_{22{\bf k}}\right) \left(E_{11{\bf k}}-E_{33{\bf k}}\right)} -
\frac{\left(E_{22{\bf k}}-\xi_{X,{\bf k+Q_Y}}\right)f\left(E_{22{\bf k}}\right)}{\left(E_{11{\bf k}}-E_{22{\bf k}}\right) \left(E_{22{\bf k}}-E_{33{\bf k}}\right)} \right. \nonumber \\
&& +\left. \frac{\left(E_{33{\bf k}}-\xi_{X,{\bf k+Q_Y}}\right)f\left(E_{33{\bf k}}\right)}{\left(E_{11{\bf k}}-E_{33{\bf k}}\right) \left(E_{22{\bf k}}-E_{33{\bf k}}\right)} \right]
\label{c4}
\end{eqnarray}
\end{widetext}
where the energies $E_{ii}$ ($i=1-3$) are the three solutions of the cubic
equation $\Delta^2 (\omega-\xi_{X,{\bf k+Q_X}}) + \Delta^2 (\omega-\xi_{Y,{\bf
k+Q_Y}}) - (\omega-\xi_{\Gamma,{\bf k}}) (\omega-\xi_{X,{\bf k+Q_X}})
(\omega-\xi_{Y,{\bf k+Q_Y}})=0$

\begin{figure}[!htb]
\centering
\includegraphics[width=.45\linewidth]{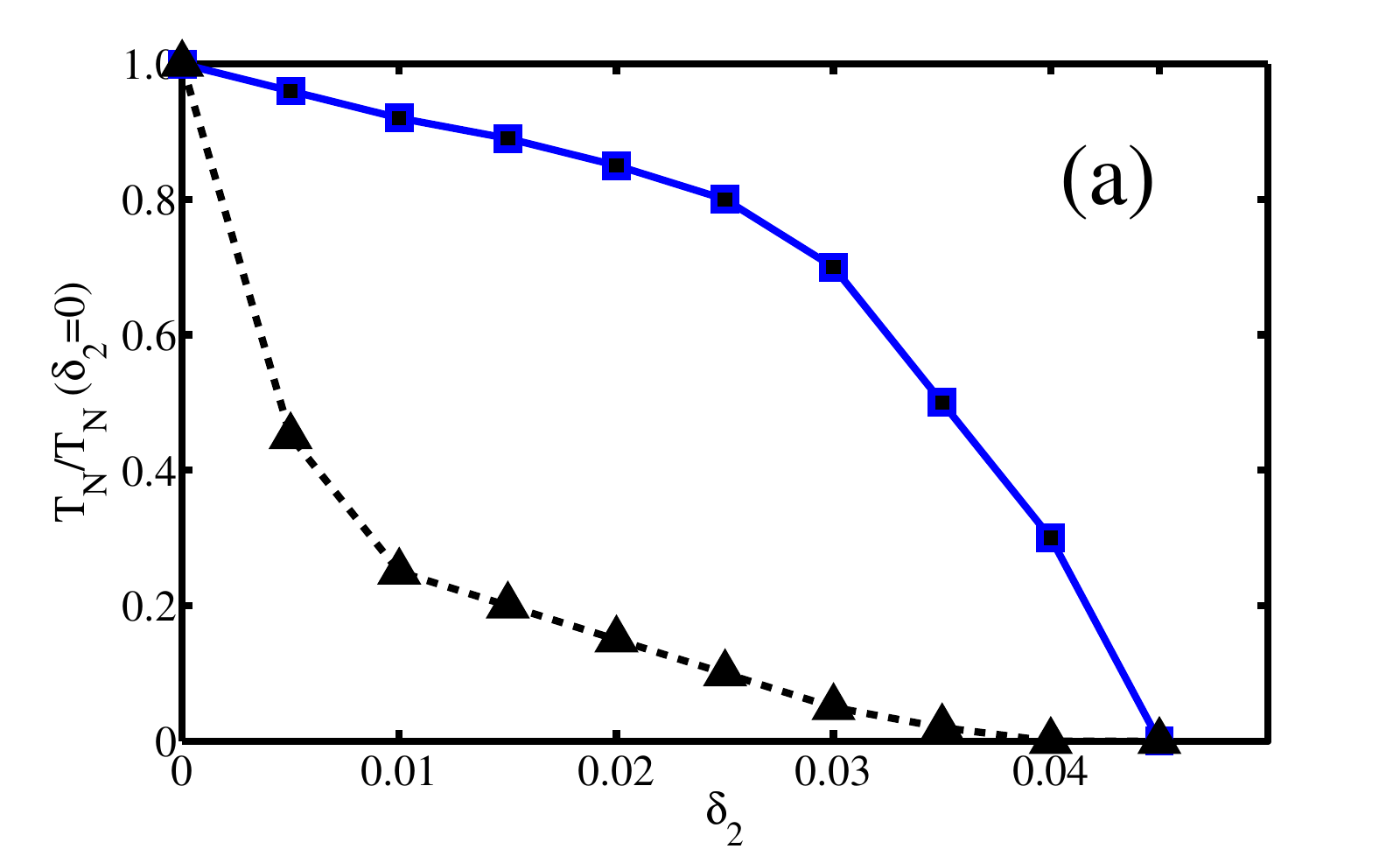}
\includegraphics[width=.45\linewidth]{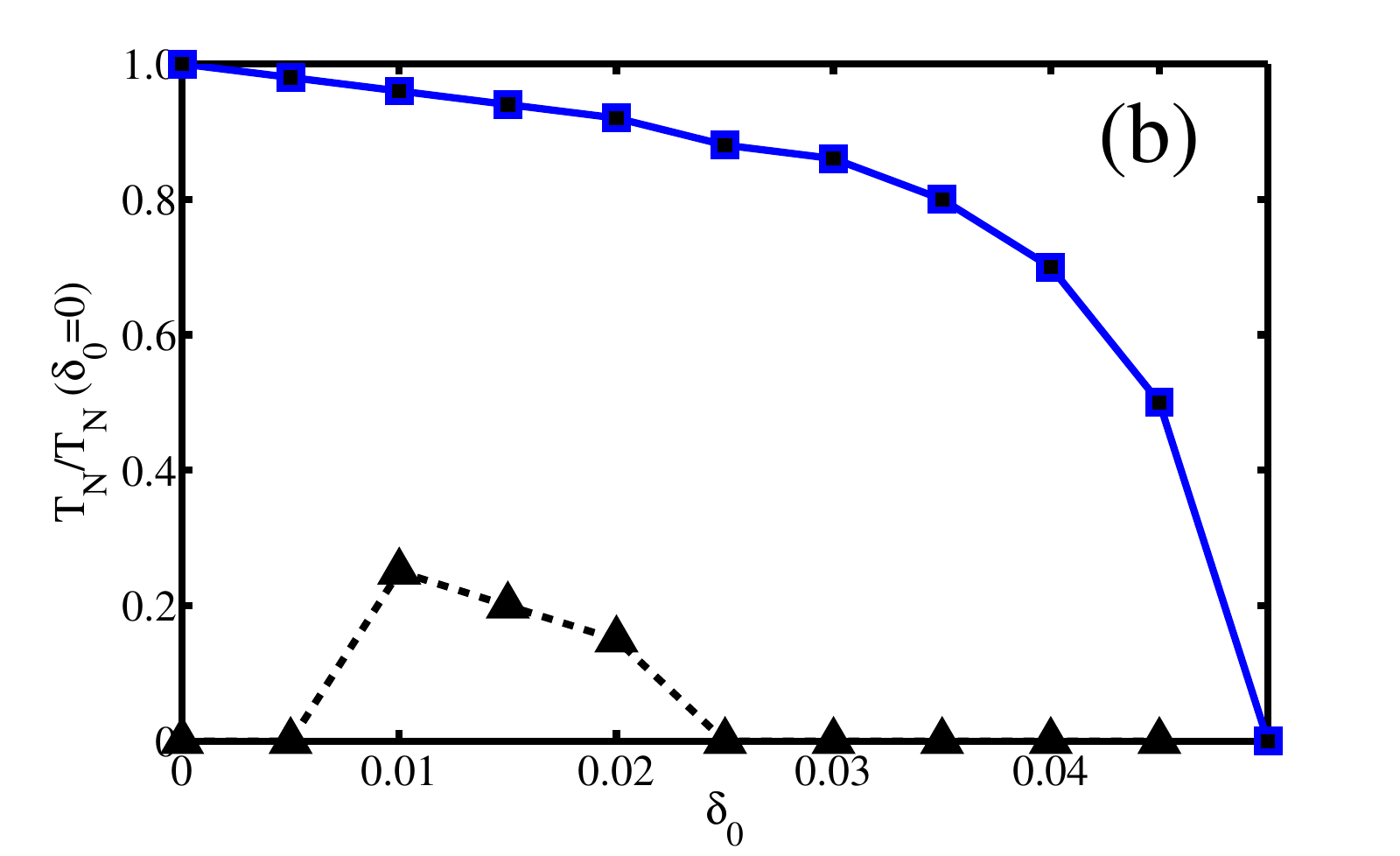}
\def\figurename{SUPPLEMENTARY FIGURE}
\caption{\footnotesize
Calculated magnetic phase diagram as a function of the
ellipticity $\delta_2$ at a finite mismatch between the electron and hole
pockets, $\delta_0=0.01$ (a) and as a function of the mismatch $\delta_0$ for
the finite ellipticity $\delta_2 = 0.01$ (b). The parameter $\delta_0$ increases
with increasing doping. The squares denote the N\'eel temperature for the stripe
phase. The triangles denote the temperature below which the $C_4$ phase wins
over the stripe phase, although the two phases remain nearly degenerate in
energy. The solid and dashed curves are guides to the eye. We used $U_{sdw}
=0.8$eV, $\epsilon_0 = 0.2$eV, and $m=1eV$. In (b), the energies
of the $C_4$ and the stripe phases are very close for all $T$, so $C_4$ phase
may actually win at low $T$ in a more generic model. } 
\label{FigT1}
\end{figure}

We solved these equations numerically together with the equation for the
chemical potential, for different values of the chemical potential mismatch
$\delta_0$ and ellipticity parameter $\delta_2$. Like in previous analysis
\cite{Eremin10,Fernandes10}, we find that the actions for $C_2$ and $C_4$ 
phases are degenerate at zero ellipticity and for equal sizes of the electron
and hole pockets ($\delta_0 =\delta_2=0$). Once ellipticity becomes non-zero,
the $C_2$ wins near the N\'eel temperature. Within the Ginzburg-Landau expansion
to order $\Delta^4_{X,Y}$, the lower free energy of the $C_2$ phase is the
consequence of the fact that the ellipticity generates the term $C
|\vec{\Delta_X}|^2|\vec{\Delta_Y}|^2$ with positive coefficient $C$, which
increases the energy of the $C_4$ phase but does not affect $C_2$ phase. Going
beyond Ginzburg-Landau approximation, we found that for small enough
ellipticity, the solution of Eq. (\ref{c4}) re-emerges below some  $T<T_N$ and,
below this $T$, $C_4$ phase becomes a local minimum. Furthermore, in some range
of $\Delta_2$, at even lower  $T<T_{\rm r}$, the free energy of the $C_4$ phase
becomes  slightly smaller than that of the $C_2$ phase, i.e., the system
undergoes a first-order transition from $C_2$ to $C_4$ magnetic phase, and
lattice $C_4$ symmetry gets restored.

At $\delta_0=0$ the region of $\delta_2$ where $C_4$ phase wins is exponentially
small. However, at a finite $\delta_0$, this region widens up. We show the
results for a particular $\delta_0$ in Supplementary Figure 3(a) and for a given
ellipticity as a function of $\delta_0$ in Supplementary Figure 3(b). A non-zero
$\delta_0$ means that the sizes of electron and hole pockets are non-equal,
i.e., that there is a  finite amount of doping.   The implication of this result
is that, at a finite doping, as the temperature is lowered, the system first
orders into a stripe $C_2$ phase in which four-fold lattice rotational symmetry
is broken ($X$ and $Y$ directions become non-equivalent),  and then, at a lower
T, it undergoes a first order transition into the $C_4$ phase in which
four-fold lattice rotational symmetry is restored.  This is consistent with our
experiment.

\end{document}